\documentclass[conference]{IEEEtran}
\IEEEoverridecommandlockouts
\usepackage{amsmath,amssymb,amsfonts}
\usepackage{algorithmic}
\usepackage{graphicx}
\usepackage{textcomp}
\usepackage{xcolor}
\usepackage{float}
\usepackage{caption}
\usepackage[
backend=biber,
style=ieee,
]{biblatex}
\usepackage[T1]{fontenc}
\usepackage[scaled=0.8]{FiraMono}
\NewDocumentCommand{\codeword}{v}{
\texttt{#1}
}
\usepackage{subcaption}

\usepackage{lastpage}
\usepackage{orcidlink}

\definecolor{lblue}{RGB}{225,232,242}
\definecolor{violette}{RGB}{246,24,225}
\usepackage{xspace}

\newcommand{\onnx}{ONNX Runtime\xspace}
\newcommand{\libtorch}{LibTorch\xspace}
\newcommand{\tflite}{TensorFlow Lite\xspace}
\newcommand{\none}{Bypass-Engine\xspace}
\newcommand{\tensorflow}{TensorFlow\xspace}
\newcommand{\pytorch}{PyTorch\xspace}
\newcommand{\cmake}{CMake\xspace}

\newcommand{\hybrid}{HNN-11k\xspace}
\newcommand{\stateful}{RNN-2k\xspace}
\newcommand{\lcnn}{CNN-29k\xspace}
\newcommand{\mcnn}{CNN-15k\xspace}
\newcommand{\scnn}{CNN-1k\xspace}
\newcommand{\macos}{MacOS\xspace}
\newcommand{\linux}{Linux\xspace}
\newcommand{\windows}{Windows\xspace}
\newcommand{\Anira}{\textit{Anira}\xspace}

\newcommand{\anira}{\textit{anira}\xspace}
\newcommand{\radsan}{RadSan\xspace}

\newcommand{\sy}{\textit{system}\xspace}
\newcommand{\sys}{\textit{systems}\xspace}

\newcommand{\ie}{\textit{engine}\xspace}
\newcommand{\ies}{\textit{engines}\xspace}

\newcommand{\ma}{\textit{model}\xspace}
\newcommand{\mas}{\textit{models}\xspace}
\newcommand{\Ma}{\textit{Model}\xspace}

\newcommand{\bs}{\textit{buffer size}\xspace}
\newcommand{\bss}{\textit{buffer sizes}\xspace}

\newcommand{\ite}{\textit{iteration}\xspace}
\newcommand{\ites}{\textit{iterations}\xspace}

\newcommand{\rep}{\textit{repetition}\xspace}
\newcommand{\reps}{\textit{repetitions}\xspace}

\newcommand{\rps}{\textit{RpS}\xspace}
\newcommand{\ri}{\textit{repetition index}\xspace}

\newcommand{\dsi}{DS-I\xspace}

\newcommand{\dsii}{DS-II\xspace}

\newcommand{\dsiii}{DS-III\xspace}
\newcommand{\lmmi}{LMM-I\xspace}
\newcommand{\lmmii}{LMM-II\xspace}
\newcommand{\lmmiii}{LMM-III\xspace}

\newcommand{\rsq}{$R^2$\xspace}
\newcommand{\rsqcon}{$R^2\textit{-conditional}$\xspace}
\newcommand{\rsqmar}{$R^2\textit{-marginal}$\xspace}
\newcommand{\rsqconbf}{$\boldsymbol{R^2}\textbf{\textit{-conditional}}$\xspace}
\newcommand{\rsqmarbf}{$\boldsymbol{R^2}\textbf{\textit{-marginal}}$\xspace}

\newcommand{\rtt}{\textit{RTT}\xspace}

\newcommand{\lmmirsqcon}{$96.0\%$\xspace}
\newcommand{\lmmirsqmar}{$95.9\%$\xspace}
\newcommand{\lmmiirsqcon}{$95.5\%$\xspace}
\newcommand{\lmmiirsqmar}{$95.4\%$\xspace}
\newcommand{\lmmiiirsqcon}{$94.9\%$\xspace}
\newcommand{\lmmiiirsqmar}{$94.9\%$\xspace}

\newcommand{\dsnonecnnmean}{$2.03\cdot 10^{-4}$}

\newcommand{\dsnonecnnse}{$3.46\cdot 10^{-6}$}
\newcommand{\dsnonecnnlci}{$1.96\cdot 10^{-4}$}
\newcommand{\dsnonecnnuci}{$2.09\cdot 10^{-4}$}

\newcommand{\dsnonehybridmean}{$3.78\cdot 10^{-4}$}

\newcommand{\dsnonehybridse}{$3.87\cdot 10^{-6}$}
\newcommand{\dsnonehybridlci}{$3.70\cdot 10^{-4}$}
\newcommand{\dsnonehybriduci}{$3.85\cdot 10^{-4}$}

\newcommand{\dsnonestatefulmean}{$5.26\cdot 10^{-5}$}

\newcommand{\dsnonestatefulse}{$1.50\cdot 10^{-6}$}
\newcommand{\dsnonestatefullci}{$4.97\cdot 10^{-5}$}
\newcommand{\dsnonestatefuluci}{$5.55\cdot 10^{-5}$}

\newif\ifmasterthesis
\masterthesisfalse

\begin{document}

\title{ANIRA: An Architecture for Neural Network Inference in Real-Time Audio Applications}

\author{\IEEEauthorblockN{Valentin Ackva \orcidlink{0009-0006-7300-6253} \textsuperscript{*}}
\IEEEauthorblockA{\textit{Audio Communication Group} \\
                  \textit{Technische Universität Berlin}\\
                  Berlin, Germany \\
                  \texttt{valentin.ackva@gmail.com}}
\and
\IEEEauthorblockN{Fares Schulz \orcidlink{0009-0003-3512-0096} \textsuperscript{*}}
\IEEEauthorblockA{\textit{Audio Communication Group} \\
                  \textit{Technische Universität Berlin}\\
                  Berlin, Germany \\
                  \texttt{fares.schulz@tu-berlin.com}}
\thanks{\textsuperscript{*}The authors have contributed equally to this work.}}

\maketitle
\begin{abstract}
Numerous tools for neural network inference are currently available, yet many do not meet the requirements of real-time audio applications. In response, we introduce \anira, an efficient cross-platform library. To ensure compatibility with a broad range of neural network architectures and frameworks, \anira supports ONNX Runtime, LibTorch, and TensorFlow Lite as backends. Each inference engine exhibits real-time violations, which \anira mitigates by decoupling the inference from the audio callback to a static thread pool. The library incorporates built-in latency management and extensive benchmarking capabilities, both crucial to ensure a continuous signal flow. Three different neural network architectures for audio effect emulation are then subjected to benchmarking across various configurations. Statistical modeling is employed to identify the influence of various factors on performance. The findings indicate that for stateless models, ONNX Runtime exhibits the lowest runtimes. For stateful models, LibTorch demonstrates the fastest performance. Our results also indicate that for certain model-engine combinations, the initial inferences take longer, particularly when these inferences exhibit a higher incidence of real-time violations.
\end{abstract}


\section{Introduction}
In recent years, neural networks have become an integral part of modern audio digital signal processing. Their applications include audio classification \cite{hershey_cnn_2017}, audio transcription \cite{bittner_lightweight_2022}, audio source separation \cite{stoller_wave-u-net_2018}, audio synthesis \cite{engel_ddsp_2020}, \cite{oord_wavenet_2016}, \cite{caillon_rave_2021} and audio effects \cite{wright_real-time_2020}. While offline processing is inherently supported, translating these architectures to real-time implementations remains challenging. To overcome this hurdle, the special requirements of real-time audio systems must be reconciled with the high computational complexity of neural network inference.\\
\indent Neural networks are commonly trained in high-level frameworks such as \tensorflow \cite{abadi_tensorflow_2016} or \pytorch \cite{paszke_pytorch_2019} in the Python programming language. While these frameworks provide a wide range of tools for training and evaluating neural networks, they are not optimized for inference and are particularly unsuitable for real-time audio applications. To circumvent this issue, one approach is to implement the network operations in high-performance programming languages such as C or C++, and then export and load the trained parameters from the high-level framework. While this approach provides the ability to optimize the inference code for the neural network being used, it is often time-consuming and lacks the flexibility to integrate different neural networks.\\
\indent To facilitate this process, a variety of inference engines have been developed in high-performance languages. These engines optimize the execution of neural networks on different hardware platforms and provide implementations of the most common neural network layers. In this paper, we focus on the three most common inference engines: \onnx, \libtorch, and \tflite. To optimize inference runtimes, these engines prioritize average processing times over real-time safety during execution. However, to ensure a continuous and uninterrupted signal flow in real-time audio applications, it is also necessary to minimize the maximum inference time or worst-case execution time.\\
\indent This need for minimum worst-case execution times, and hence deterministic runtimes in audio callbacks, has led to the adoption of certain real-time principles for conventional audio DSP algorithms. These principles include the avoidance of system calls such as dynamic memory allocation, unconditional thread locks, interaction with the thread scheduler, or code awaiting hardware on operating systems with non-real-time-safe architectures, as they may exhibit undeterministic runtimes \cite{bencina_interfacing_2014}. For the aforementioned inference engines, tests have been conducted to verify their adherence to these real-time principles \cite{chowdhury_rtneural_2021}, \cite{stefani_comparison_2022}. However, different conclusions have been derived. Our work validates the assumptions made in \cite{chowdhury_rtneural_2021} and quantifies the real-time violations observed in inference executions.\\
\indent To ensure that these inference engines can be executed safely in real-time audio applications, we propose a library called \anira. The library provides an architecture to outsource the processing of the inference engines to a thread pool, allowing the audio callback to remain free of blocking operations. Furthermore, the library handles processing of the inference engines in discrete chunks, a critical feature for inference in neural networks that are designed to process inputs of a predetermined size. Ultimately, the library aims to provide a framework for integrating neural networks into real-time audio applications. It is currently compatible with the three inference engines \onnx, \libtorch, and \tflite, the operating systems \linux, \macos, and \windows, and supports various types of neural networks. The source code of the library is publicly available \cite{schulz_anira_nodate}.\\
\indent \Anira's \linux compatibility allows it to support not only desktop applications but also multicore embedded systems and servers. While there exist pipelines that provide insight into the deployment of neural networks in specific embedded systems \cite{stefani_real-time_2023}, \cite{pelinski_pipeline_2023}, these pipelines are limited in their generalizability. In contrast, \anira employs a library design that aims at an abstraction on a higher level by offering an API compatible with multiple configurations. Moreover, the library's ability to manage inference tasks across multiple threads allows for the optimal utilization of the underlying hardware's processing capabilities. The library's integration with the \cmake build system \cite{kitware_inc_cmake_nodate} enables its incorporation into a wide range of build pipelines and environments.\\
\indent In addition to its real-time implementation, the computational load of a neural network is another important factor for real-time executability. While several neural networks have been evaluated for real-time performance, the benchmarks often lack comparability due to different inference implementations and the use of different inference engines or programming languages \cite{caillon_rave_2021}, \cite{wright_real-time_2020}, \cite{steinmetz_efficient_2022}. Further, the missing comprehensive documentation of the benchmarking procedure make it difficult to understand the circumstances under which the benchmarks were conducted.\\
\indent We address this issue by including built-in benchmarking capabilities for evaluating the real-time performance of neural networks in our library. The benchmarks may be executed with a variety of configurations, including different buffer sizes, which is a standard feature in most real-time audio environments. The library's compatibility with a range of inference engines, neural networks, and operating systems allows for the execution of benchmarks under identical conditions, thereby ensuring comparability and reproducibility.\\
\indent To date, the neural network runtimes for the three inference engines in the audio domain have only been compared for audio classifier models consisting of dense and batch normalization layers \cite{stefani_comparison_2022}. In this work, the execution times of three neural network models for audio effect emulation were measured under various circumstances. Each model incorporates a distinct set of layers, including convolutional, recurrent, and stateful layers. Statistical models of the resulting datasets reveal that certain inference engines are more suited to certain neural network layers than others, and that in some cases, the initial inferences performed by the engines are slower than subsequent inferences. In a follow-up study, we examine the performance of the inference engines on a range of differently sized convolutional neural networks. Our findings demonstrate that the engines exhibit varying levels of efficiency across different network sizes.\\
\indent The remainder of this paper is organized as follows: Section \ref{sec:implementation} describes the implementation of the library, including its built-in benchmarking tools. Section \ref{sec:methods} details the methods used to verify the real-time safety of the inference engines and the library as well as evaluate their performances. The results are then presented in Section \ref{sec:results} and discussed in Section \ref{sec:discussion}.
\section{Implementation}\label{sec:implementation}
\subsection{Inference Engines}\label{sec:ie}
The \anira library enables the use of various inference engines as backends. At present, \tflite 2.16.1 \cite{tensorflow_tensorflow_nodate}, \onnx 1.17.1 \cite{microsoft_onnx_nodate}, and \libtorch 2.2.2 \cite{pytorch_libtorch_nodate} are supported. The choice of these engines was guided by the following criteria: compatibility with C/C++, cross-platform operation, support for common processor architectures, open-source licensing favorable for commercial use, and support of common neural layers. All inference engines are configured with the default settings, with the exception of the number of threads that the engines can utilize, which is limited to one, because \anira manages the parallelization of the inference tasks.
\subsection{Interface}
\begin{figure*}[h]
    \centering
    \includegraphics[width=\textwidth]{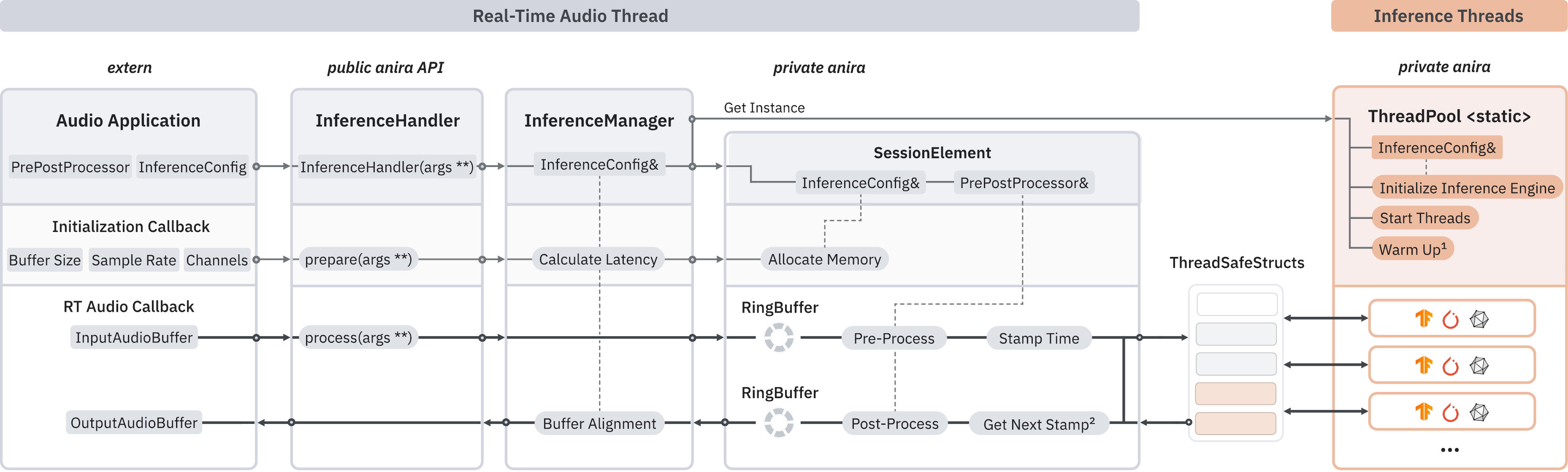}
    \caption{Architectural overview of the \anira library. The diagram illustrates the design and the principle components involved in the inference process. The overview is divided into three main sections: the Constructor, the Initialization Callback, and the Real-Time Audio Callback. The "\textit{Warm Up}" operation, marked with $^1$, is optional and allows for a set number of inferences before the start of the audio callback. The "\textit{Get Next Stamp}" operation, marked with $^2$, can also be configured to await a specified period of time for the processed data to become available.}
    \label{fig:anira-architecture}
\end{figure*}

The interface is designed to streamline real-time inference processes across a wide range of neural network types. Internally, \anira then manages all critical operations, including thread allocation, memory handling, and execution control. Figure \ref{fig:anira-architecture} provides an architectural schematic of the library, illustrating the external audio application (\textit{extern}), the interface of the library (\textit{public} \anira \textit{API}), and the internal components (\textit{private} \anira).

\indent For interaction with the library, the \codeword{InferenceHandler} class is provided. This class serves as the interface for configuring and executing neural network models within the \anira framework. Its constructor requires an \codeword{InferenceConfig} struct to pass the inference settings as outlined in Table \ref{tab:anira-config}. In addition, an optional \codeword{PrePostProcessor} can be configured to perform model-dependent adjustments on the input and output data, such as extending the input sequence with past audio data to meet the specific requirements of the model. It is noteworthy that resampling, which is often necessary due to the sample rate dependency of many neural networks \cite{carson_sample_2024}, is currently not supported within \anira and must, therefore, be handled externally.
\begin{table} [h]
\centering
\caption{Parameters for the \texttt{InferenceConfig}, used to configure the \anira library. All options, with the exception of those with a default value, are mandatory. Options denoted with \textsuperscript{$\dagger$} are dependent on the enabled inference engines within \anira. Each engine requires its own set of these parameters.} \label{tab:anira-config}
\begin{tabular}{|l|l|}
\hline
\textbf{Option} & \textbf{Description} \\  
\hline
model\textsuperscript{$\dagger$} & binary / path of model ($\ast$.pt, $\ast$.onnx, $\ast$.tflite) \\
model\_input\_shape\textsuperscript{$\dagger$} & input shape, e.g. \{batch\_size, input\_size, 1\} \\ 
model\_output\_shape\textsuperscript{$\dagger$} & output shape, e.g. \{batch\_size, output\_size, 1\} \\
max\_inference\_time & maximum measured inference time (ms) \\
model\_latency & internal latency of the model (samples) \\
warm\_up & number of warm up inferences (default = 0) \\
wait\_in\_process\_block & proportional wait time to receive data \\
 & (value between 0.0 - 0.95, default = 0.0) \\
bind\_session\_to\_thread & bind instance to one thread (default = off) \\
num\_threads & number of threads created by thread pool \\
 & (default = number of concurrent threads - 1) \\
\hline
\end{tabular}
\end{table}

\indent The initialization of the \codeword{InferenceHandler} instance requires the specification of the buffer size, the number of channels, and the sample rate of the host. Once initialized, the necessary memory allocations are completed, ensuring that \anira operates safely within the audio callbacks. During runtime, the inference engine can be selected. The integration of the library is demonstrated through a JUCE audio plugin example included in the repository \cite{schulz_anira_nodate}.
\subsection{Internal Architecture}\label{sec:internal}
Besides the \textit{public} \anira \textit{API}, the library consists of three main internal components: the \codeword{InferenceManager}, the \codeword{SessionElement}, and the \codeword{ThreadPool}. The interplay of these three classes enables the separation of the inference process from the audio callback onto independent high-priority threads to ensure real-time safety. This also permits the processing of audio data in discrete chunks independent of the host buffer size, which is crucial for models with a fixed input shape. In this process, the \codeword{InferenceManager} is responsible for buffer alignment and latency calculation, the \codeword{SessionElement} stores the audio data and inference results, and the static \codeword{ThreadPool} orchestrates the inference threads.\\
\indent As data sharing between threads requires precautions to prevent race conditions and deadlocks, the \codeword{SessionElement} provides specific \codeword{ThreadSafeStucts}. \Anira offers two different implementations of these structures to regulate access to the audio data and to submit inference tasks to the \codeword{ThreadPool}\nolinebreak. The first option follows strict real-time safety rules with raw atomic values, while the second option uses semaphores to allow a controlled blocking operation that can further reduce latency (cf. Section \ref{sec:latency}). \codeword{std::atomic<bool>}\nolinebreak /\nolinebreak \codeword{std::binary_semaphore} ensure that the audio data can only be accessed by one thread, while \codeword{std::atomic<int>}\nolinebreak /\nolinebreak \codeword{std::counting_semaphore} are used to submit inference tasks to the queue. Timestamps, realized through inference buffer counting, ensure that the processed audio output is in the correct sequence.\\
\indent The static design choice of the \codeword{ThreadPool} is intentional to overcome a problem called oversubscription that may arise when multiple instances are created, each with its own high-priority inference thread. This becomes problematic when the demand for real-time threads exceeds the number of available hardware threads, leading to a degradation in overall performance \cite{williams_managing_2012}. The maximum number of threads in the pool is configurable. For stateless models, all inference threads are created upon initialization. This approach enables accelerated inference through parallel inferencing.
\subsection{Latency}\label{sec:latency}
To maintain a stable and uninterrupted signal flow, the \codeword{InferenceManager} determines the minimum latency that must be applied. Besides the inherent latency of the host, defined by its host buffer size, there are three different key factors that contribute to the overall latency: potential buffer size mismatches between the host buffer size and the model input size, the worst-case inference time, and the internal latency of the model. Whereas the host buffer size, model input size, and internal model latency are distinct values, the worst-case inference time can only be estimated by extensive benchmarks (cf. Section \ref{sec:benchmarks}). This is due to the non-deterministic runtimes of the inference engines, which are a consequence of the real-time violations that the engines exhibit (cf. Section \ref{sec:results-radsan}).\\
\indent To account for the potential buffer size mismatch, we rely on the algorithm provided by \citeauthor{letz_callback_2001} \cite{letz_callback_2001}. In the paper, \citeauthor{letz_callback_2001} identified the minimum latency required to match buffers of differing sizes $A$ and $B$. This was achieved by analyzing the remainder of the division of integer multiples $n$ of $A$ by $B$, $\mod(nA,B)$, until the pattern repeats.\\
\indent Furthermore, it is noteworthy that the buffer size mismatch may also incorporate a varying number of inferences that can be performed within one audio callback. We leverage the pattern approach proposed by \citeauthor{letz_callback_2001} to determine the maximum number of possible inferences that can be expected within an audio callback and derive the combined worst-case inference time. However, since the request for inferenced data to the thread pool is made only once per audio callback, the worst-case inference time must be rounded up to the next integer multiple of the host buffer size.\\
\indent Finally the model's internal latency is added to the total latency calculation. The resulting formula is as follows: 
\begin{equation}
    L_{\text {total }}= H_{\text{adapt}} + \left\lceil\frac{I_{\text{max}}}{H_{\text{host}}}\right\rceil \cdot H_{\text{host}} + M_{\text{int}} \label{eq:latency}
\end{equation}
\noindent The term $H_{\text{adapt}}$ represents the latency required to adapt the buffer sizes, $I_{\text{max}}$ denotes the combined worst-case inference time in samples, $H_{\text{host}}$ is the host buffer size, and $M_{\text{int}}$ refers to the internal latency of the model.\\
\indent Given that the value of the second term in (\ref{eq:latency}) is allways greater than or equal to $H_{\text{host}}$, $L_{\text{total}}$ will never reach zero. For some applications, this may be an unacceptable outcome. Therefore, \anira includes an option within the \codeword{InferenceConfig} to specify a wait\nolinebreak\space time and thus introduce a controlled blocking operation to the audio callbacks to receive processed data (cf. Table \ref{tab:anira-config}). The blocking operation is implemented as a \codeword{std::counting_semaphore::try_acquire_until} method, and requires the usage of the \codeword{ThreadSafeStucts} implementation with semaphores (cf. Section \ref{sec:internal}). This enables \anira to achieve real-time processing with no additional latency introduced by the library, for certain configurations. Subsequently, (\ref{eq:latency}) is modified to:
\begin{equation}
    L_{\text {total }}= H_{\text{adapt}} + \left\lceil\frac{I_{\text{max}} - H_{\text{host}} \cdot W}{H_{\text{host}}}\right\rceil \cdot H_{\text{host}} + M_{\text{int}} \label{eq:latency-wait}
\end{equation}
\noindent In this context, $W$ denotes the proportional wait time, as defined in the \codeword{InferenceConfig}.
\subsection{Benchmarks}\label{sec:benchmarks}
The benchmarks are designed to evaluate the buffer execution runtimes of the \anira library across a range of different configuration combinations. These configuration options include the selected inference engine, the neural network model, and the buffer size. Technically, the benchmarking functionality is implemented within the Google Benchmark \cite{google_llc_google_nodate} and Google Test \cite{google_llc_google_nodate-1} framework. When building the library, the benchmarking functionality can be enabled optionally. Subsequently, the benchmarks can be executed as unit tests.\\
\indent In order to ensure that the runtime results are representative of the processing times that occur in real-time audio applications, it is essential that the test environment closely simulates the target environment. This is achieved by implementing a series of steps typically found in audio applications. Firstly, the priority of the benchmark process is set to the highest possible level. Subsequently, a static instance of the \codeword{InferenceHandler} class is constructed and initialized. In the next step, the time required for an input buffer, filled with random generated samples, to be processed by the \codeword{InferenceHandler} instance is measured. This process is repeated a defined number of times, with each iteration being measured individually in order to simulate the continuous processing of audio buffers in a real-time audio environment. Once the iterations have been completed, the \codeword{InferenceHandler} instance is destroyed, and the benchmark process sleeps for as long as it took to process all the iterations. The aforementioned steps are then repeated for a specified number of repetitions.\\
\indent This methodology allows for the simulation of multiple repetitions of the initial inference runs, and therefore, the warm-up phases of the inference engines. Optinally this procedure can be repeated for different configurations.
\section{Methods} \label{sec:methods}
\subsection{Selected Neural Networks} \label{sec:nn}
To evaluate the inference engines and the \anira library, three neural network models are selected. These models are tailored for end-to-end audio processing tasks, with a special focus on real-time audio effects emulation. Each of these models represents one of the prominent neural network architectures that have been identified as particularly suitable in recent research: convolutional neural networks (CNN), recurrent neural networks (RNN), and hybrid configurations \cite{vanhatalo_review_2022}. The hyperparameters of the models are selected to ensure efficient real-time performance. All models have been trained with the same dataset, a three minute dry and wet recording of an Ibanez TS9 Tube Screamer guitar pedal from \cite{keith_bloemer_guitarlstm_nodate}.
\subsubsection{Convolutional Neural Network}
\begin{figure}[h]
\centering
\includegraphics[width=0.418\textwidth]{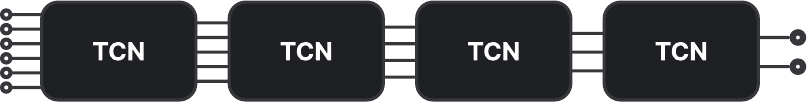}\\
\smallskip
\includegraphics[width=0.242\textwidth]{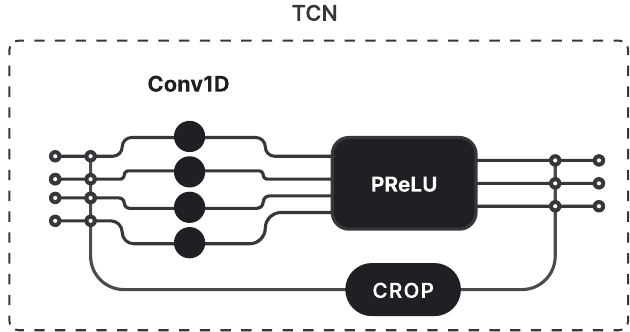}
\caption{Schematical representation of the selected CNN model architecture. The top figure shows the overall model structure, while the bottom figure provides a detailed view of the TCN block.}\label{fig:cnn}
\end{figure}

The chosen CNN is a Temporal Convolutional Network (TCN) inspired by the work of \citeauthor{steinmetz_efficient_2022} \cite{steinmetz_efficient_2022}. TCN architectures use dilated causal convolution layers to capture long-range dependencies in sequential data, achieving a wide receptive field without the need for a deep layered structure. The dilation grows with each TCN block, so that the dilation of the convolution layer in the $i$-th TCN block is equal to $d_i = d^{i-1}$, where $d$ denotes the dilation factor. In contrast to the model architecture presented by \citeauthor{steinmetz_efficient_2022}, the FiLM and batch normalization layers have been removed for simplicity and to focus on streamlined, efficient computation. As a result, the model consists of TCN blocks containing a causal convolution layer, a PReLU activation layer, and a residual connection (cf. Fig. \nolinebreak\ref{fig:cnn}). Three different TCN models are defined with different hyperparameter combinations in Table \ref{tab:nn-hyperparameters}: a large (\lcnn), medium (\mcnn), and small model (\scnn).
\subsubsection{Recurrent Neural Network}
\begin{figure}[h]
\centering
\includegraphics[width=0.297\textwidth]{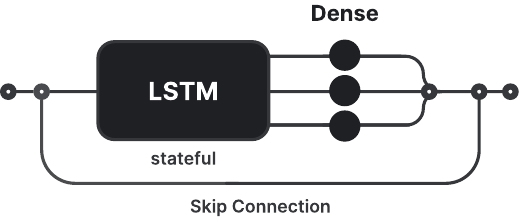}
\caption{Schematical representation of the selected RNN model architecture.} \label{fig:rnn}
\end{figure}

The chosen RNN is adapted from the work of \citeauthor{wright_real-time_2020} \cite{wright_real-time_2020}. The model uses a stateful LSTM layer, a dense layer, and a residual connection to predict one audio sample based on the previous samples (cf. Fig. \ref{fig:rnn}). It is designed to process audio samples sequentially, with the LSTM layer retaining information across inputs. This means that multithreading is not supported, as the internal state of the LSTM layer must be updated after each buffer. Also, this network is only compatible with \libtorch and \tflite, as \onnx inherently supports only stateless operations \cite{noauthor_onnx_2023}. However, it should be noted that the state of the model could be handled externally, but, as this would alter the model graph, we have chosen to exclude \onnx from the RNN comparison. At the end of the model architecture, the dense layer processes the output of the LSTM sequence and outputs the resulting sequence of ouput samples. The selection of the hyperparameters for the defined model, \stateful, is detailed in Table \ref{tab:nn-hyperparameters}.
\subsubsection{Hybrid Neural Network}
The hybrid configuration model is a stateless LSTM layer combined with two convolutional layers and a dense layer \cite{keith_bloemer_guitarlstm_nodate}, \cite{keith_bloemer_neural_2021}. Stateless LSTM layers are a variant of LSTM layers that does not retain information across batches. For capturing information from past signals but limiting the size of the sequence that gets processed by the LSTM layer, the network incorporates two 1-D convolutional layers with a stride greater than one before the LSTM layer. This leads to a down-sampling of the input sequence, allowing the network to abstract features from a sequence of audio samples. At the end of the model architecture the dense layer processes the output of the last LSTM cell and outputs a single audio sample. The hyperparameters selected for the defined model, \hybrid, are detailed in Table \ref{tab:nn-hyperparameters}.
\begin{figure}[h]
\centering
\includegraphics[width=0.33\textwidth]{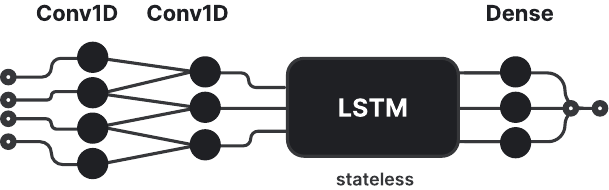}
\caption{Schematical representation of the selected hybrid model architecture.} \label{fig:hybrid}
\end{figure}

\begin{table} [h]
\centering
\caption{Hyperparameters for CNN, RNN, and hybrid models. The variables $b$, $k$, $c$, and $d$ represent the number of TCN blocks, kernel size, number of convolution channels, and dilation factor, respectively, for the CNN models. For the RNN model, $h$ represents the number of hidden units in the LSTM layer. In the hybrid model, $s$ and $c$ denote the stride and number of channels in the convolutional layers, while $h$ represents the number of hidden units in the LSTM layer. The parameter count for each model is also provided. Due to differences in the implementation of the LSTM layer, the parameter number differs for \tensorflow and \libtorch, denoted by * and **, respectively.} \label{tab:nn-hyperparameters}
\begin{tabular}{|c|c|c|c|c|c|c|c|}
\hline
\textbf{Model} & $\boldsymbol{b}$ & $\boldsymbol{k}$ & $\boldsymbol{c}$ & $\boldsymbol{d}$ & $\boldsymbol{s}$ & $\boldsymbol{h}$ & \textbf{Parameters} \\  
\hline
\lcnn & 4 & 13 & 32 & 10 & - & - & 29669 \\
\mcnn & 3 & 13 & 32 & 10 & - & - & 15300 \\
\scnn & 2 & 13 & 32 & 10 & - & - & 931 \\
\stateful & - & - & - & - & - & 20 & 1781* / 1861** \\
\hybrid & - & - & 16 & - & 12 & 36 & 10965* / 11109** \\
\hline
\end{tabular}
\end{table}
\indent All neural network models in this work were implemented and trained using both major deep learning frameworks, \pytorch and \tensorflow. For each model, an existing implementation in one framework was altered to better fit the use cases of this work. Subsequently, an equivalent model was implemented for the alternative framework. These altered models, along with their equivalent implementations in the alternate framework, are available online \cite{schulz_anira_nodate}. Following this, the \tensorflow models were exported to the \tflite format, while the \pytorch models were exported to the \libtorch and \onnx formats. The \tensorflow models were not exported to the \onnx format due to the absence of official support for this operation.
\subsection{Real-Time Evaluation} \label{sec:real-time-eval}
To assess whether the inference engines and \anira conform to real-time principles, we utilize a code sanitizer to identify potential violations. Code sanitizers function by embedding instrumentation into the program’s binary, enabling, for example, the monitoring of memory operations \cite{serebryany_addresssanitizer_2012}. In this work we are leveraging the Real-time Sanitizer (\radsan) \cite{trevelyan_realtime_nodate}, which is tailored to detect common real-time principle violations like memory allocation, deallocation, and thread synchronization. \\
\indent The test was conducted on a \linux x64 system to determine the quantity and types of real-time violations. For the three inference engines, the inference execution of the models described in Section \ref{sec:nn} was monitored over 50 consecutive inferences. To validate the real-time safe operation of \anira, the library's process method was monitored across different buffer sizes, models and selected inference engines.   
\subsection{Operationalization and Datasets} \label{sec:datasets}
The datasets utilized for the statistical analysis were compiled from a set of benchmarks (cf. Section \ref{sec:benchmarks}) that were executed on a variety of operating systems and hardware platforms (\sys). The first \sy is a MacBook Pro with an Intel Core i9-9980HK processor and 32 GB of RAM, running Arch \linux, kernel version 6.8.4. The second \sy is a MacBook Pro with a M1 processor and 16 GB of RAM, running \macos 14.4.1. The third \sy is a HP ZBook Fury 16 G9, equipped with an Intel Core i7-12800HX processor and 32 GB of RAM. It runs the \windows 11 operating system. All datasets consist of runtime measurements for processing 50 consecutive buffers, indexed as \ites, for specific configurations, repeated identically 10 times (\reps). The measured runtimes are divided by the buffer size in samples to obtain the runtimes per sample (\rps).\\
\indent Configurations differ in the \sy and the inference engine (\ie), which is either one of the \ies from Section \ref{sec:ie} or a \none. The latter implements the same pre- and post-processing stages as the other \ies, but writes the input samples directly to the output samples in the inference stage. Other configuration options entail the neural network models as defined in Section \ref{sec:nn}. Although the exported neural network models for \libtorch, \onnx, and \tflite are distinct files with disparate file formats ($\ast$.pt, $\ast$.onnx and $\ast$.tflite), the underlying architectures are identical across all three \ies. For the purposes of statistical analysis, the neural network models have been grouped into categories of similar \mas. The categories are defined in the Table \ref{tab:nn-hyperparameters}. Finally, all configurations were benchmarked with \bss ranging from 64 to 8192 samples. Despite the \anira library's capacity to process host buffers, which may have varying dimensions from those required by the inferred \ma, the benchmarks utilized \mas that precisely aligned with the selected \bss in the benchmark. This approach was employed to permit an evaluation of the influence of the number of samples inferred simultaneously on the relative time taken to infer one sample.\\
\indent From these observations, two datasets were created for the main study to ensure complete combinations. Since \onnx does not support stateful operations \cite{noauthor_onnx_2023}, the \stateful is excluded in the \dsi. Furthermore, only one of the three CNN \mas is included in the \dsi, namely the \lcnn. The \dsi contains a total of 96,000 \rps observations ($N = 96,000$), representing the bottom layer of a seven-layer data structure -- $3\times4\times2\times8\times10\times50$. The initial four layers represent the variables \sy, \ie, \ma, and \bs, while the fifth and sixth layers correspond to the \rep and \ite respectively ($\sy\times\ie\times\ma\times\bs\times\rep\times\ite$). The \dsii is analogous to the \dsi, but it incorporates the \stateful and excludes \onnx. This results in $N = 108,000$ and the form $3\times3\times3\times8\times10\times50$.\\
\indent For the follow-up study, which compares the \ie differences over different hyperparameter combinations of the CNN architecture, the \dsiii was created. The \dsiii contains only the CNN \mas (\scnn, \mcnn, \lcnn) and excludes the \none. It has the form $3\times3\times3\times8\times10\times50$ with $N = 108,000$. All measurements were conducted on the \anira version 0.1.0 and all datasets are publicly available \cite{schulz_anira-benchmark-evaluation_nodate}.
\subsection{Library Performance} \label{sec:met:libperf}
For evaluating the runtime performance of the \anira library, the \dsii is employed, as it incorporates observations of the \none for the \lcnn, \hybrid and \stateful. For each of these \mas, a series of descriptive statistics were calculated using the pastecs package \cite{grosjean_pastecs_nodate} in the programming language R \cite{r_core_team_r_nodate}. This process involved filtering the dataset for the \none category in the \ie variable.
\subsection{Statistical Modeling} \label{sec:stat}
In order to compare the performance of the \ies, and in particular to ascertain the effect of the \ite and the \bs, we use statistical data analysis to find the best predictors for \rps variations obtained in our datasets. Due to the multilevel structure of the datasets, linear mixed effects models (LMMs) are employed for data analysis.\\
\indent In these models, the \rps measure is the dependent variable and the \sy, \ie, \ma, \bs, and \ite are fixed effects. All independent variables are considered factors, including the \bs and \ite, since their influence on the \rps measures is not necessarily deemed to be linear. Besides the fixed effects, all possible interactions between fixed effects are modeled. To account for \rps fluctuations that may arise due to the process scheduling of the operating systems, each \rep is given a unique identifier independent of the chosen configuration. The \ri, therefore, represents a specific time window and is included in the LMMs as a random intercept. Since the benchmark process sleeps after every \rep (cf. Section \ref{sec:benchmarks}) the correlation between the time windows is assumed to be negligible.\\
\indent For each dataset, a distinct LMM was fitted using the lme4 package \cite{bates_fitting_2015} in R with maximum likelihood estimation. This results in the statistical models \lmmi for \dsi, \lmmii for \dsii, and \lmmiii for \dsiii, all of which underwent an analysis of variance (ANOVA) to determine the significance of the fixed effects and their interactions.\\
\indent Subsequently, the estimated marginal means for selected predictor combinations, including their standard errors (SE) and confidence intervals (CI), were determined with the emmeans package \cite{russell_v_lenth_emmeans_nodate}, as well as various post-hoc tests (cf. Table \ref{tab:post-hoc-tests-main}). These include pairwise comparisons to identify significant differences between factor combinations, while accounting for all other variables in the statistical model. To identify significant outliers in the \ite predictor, effect contrasts were employed. All $p$-values are adjusted for multiple comparisons using the Bonferroni-Holm method, as outlined in \cite{sture_holm_simple_1979}. Given the size of the datasets, with $N \approx 100,000$, only highly significant p-values are considered in all tests, with a significance level of $p < 0.0001$. 
\begin{table} [h]
\centering
\caption{Overview of conducted post-hoc tests on the three different LMMs.} \label{tab:post-hoc-tests-main}
\begin{tabular}{|c|c|c|c|}
\hline
\textbf{Statistical model} & \textbf{Predictor Combinations} & \textbf{Type} \\ 
\hline
LMM-I, II & \ie & pairwise \\
LMM-I, II \& III & \ie, \ma & pairwise \\
LMM-I \& II & \ite, \ie, \ma & effect \\
LMM-I \& II & \bs, \ie & pairwise \\
\hline
\end{tabular}
\end{table}
\section{Results} \label{sec:results}
\subsection{Real-Time Safety Tests} \label{sec:results-radsan}
\begin{figure}[h]
    \centering
    \includegraphics[width=0.49\textwidth]{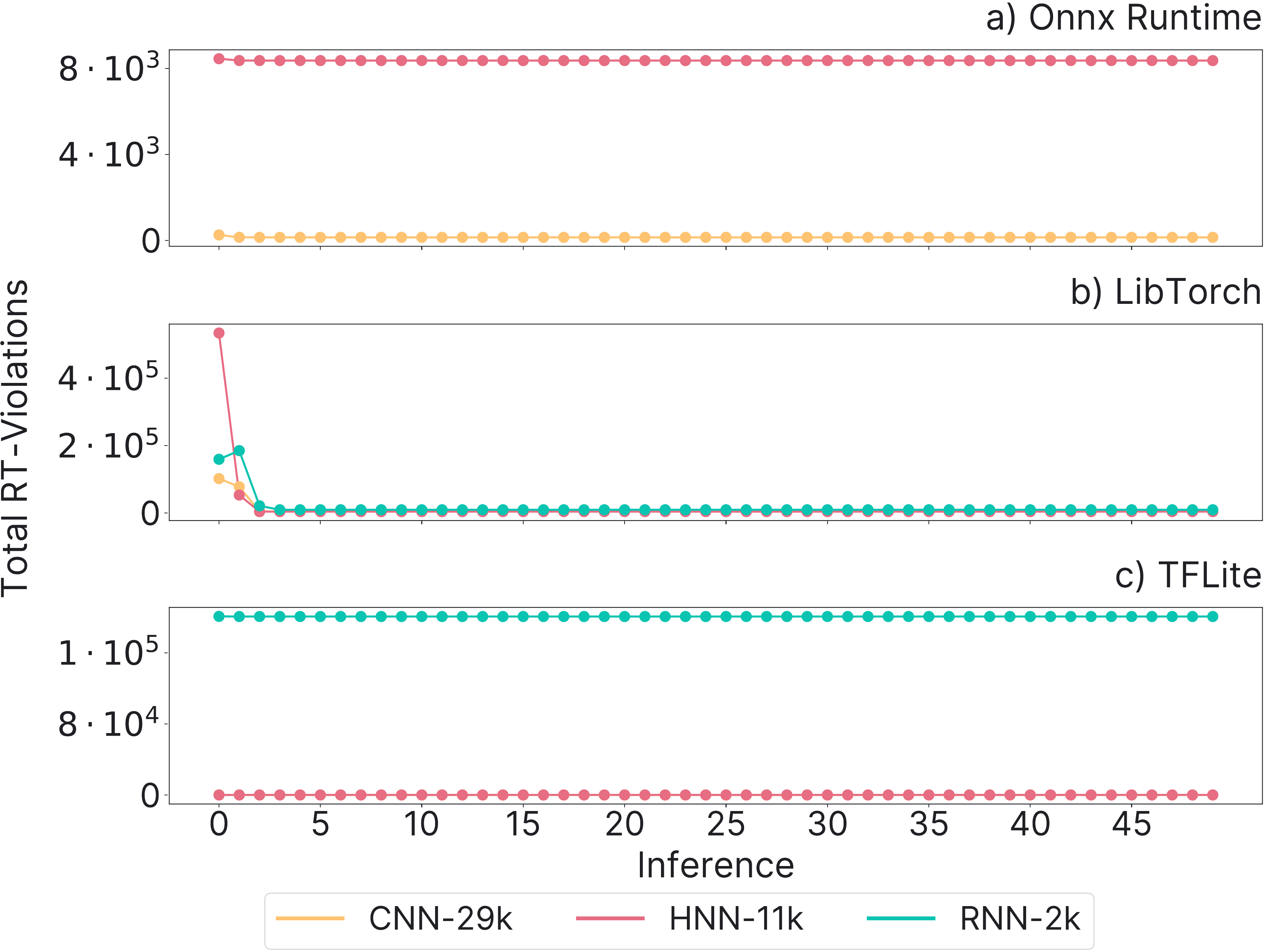}
    \caption{Total number of real-time violations detected by \radsan for each inference engine, neural network model, and inference count. It is noteworthy that none of the presented graphs reach zero.}
    \label{fig:RadSan}
\end{figure}

The real-time safety tests demonstrate a persistent pattern of real-time violations across all inference engines, neural network models, and inference counts. However, the frequency of these violations varies, as illustrated in Fig. \ref{fig:RadSan}.
\ifmasterthesis
\else
\\
\fi
\indent \libtorch exhibites the highest number of real-time violations, particularly during the initial inferences. Subsequent inferences also show a consistently higher number of violations compared to the other inference engines. In each inference cycle for each model, intensive non-real-time-safe memory operations, such as \codeword{malloc}, \codeword{free}, and \codeword{calloc}, as well as thread synchronization operations using mechanisms like \codeword{pthread_mutex_lock} and \codeword{pthread_rwlock_rdlock}\nolinebreak, are involved. Additionally, the \lcnn model performs the operation \codeword{sleep} and utilises file access functions, such as \codeword{fopen}, during the initial inference. In contrast, \tflite's real-time violations are exclusively related to memory operations, specifically \codeword{malloc}, \codeword{free} and \codeword{aligned_alloc}\nolinebreak. While the \hybrid and \lcnn models demonstrate minimal memory activity, the \stateful model exhibites a consistently high number of violations across all inference stages. For \onnx, the \hybrid model consistently shows intensive memory usage, while the \lcnn model exhibits notably fewer violations. The real-time violations observed in \onnx include \codeword{malloc}, \codeword{free} and \codeword{posix_memalign} operations. Extensive testing of the \anira library has revealed no violations.
\subsection{Library Performance}
In order to assess the runtime performance of the \anira library across a range of \mas, and consequently different pre- and post-processing steps, descriptive statistics are depicted in Table \ref{tab:library-performance}.
\begin{table} [h]
\centering
\caption{Descriptive statistics of \rps observations for the \none across different \mas. The CI employed is 95\%. All values are expressed in milliseconds per sample.} \label{tab:library-performance}
\begin{tabular}{|c|c|c|c|c|}
\hline
\textbf{\ma} & \textbf{Mean} & \textbf{SE} & \textbf{CI lower\text{ }/\text{ }upper} \\
\hline
\lcnn & \dsnonecnnmean & \dsnonecnnse & \dsnonecnnlci\text{ }/\text{ }\dsnonecnnuci \\
\hybrid & \dsnonehybridmean & \dsnonehybridse & \dsnonehybridlci\text{ }/\text{ }\dsnonehybriduci \\
\stateful & \dsnonestatefulmean & \dsnonestatefulse & \dsnonestatefullci\text{ }/\text{ }\dsnonestatefuluci \\
\hline
\end{tabular}
\end{table}

\indent The results indicate that the \stateful exhibits the lowest mean \rps value across all \mas, with a mean of \xspace\dsnonestatefulmean\xspace milliseconds per sample. This is to be expected, as the \stateful requires the least pre- and postprocessing. The values can be compared with the time per sample that an audio application has to process a buffer at a given sample rate. This real-time threshold per sample (\rtt) is equal to the inverse of the sample rate ($\rtt = 1/\text{sample rate}$).\\
\indent For a sample rate of 48 kilohertz, this results in an \rtt of 0.0208 milliseconds. Consequently, the processing time of the \anira library without an actual inference stage is substantially lower than the \rtt for all \mas.
\subsection{Statistical Models} \label{sec:eval-results}
As a preliminary step in evaluating the significance of the statistical models, the coefficients of determination are examined. The \rsq-values presented in Table \ref{tab:rsqrvalues} indicate that in all three LMMs, the fixed effects and their interactions explain a substantial portion of the variance in the \rps observations. Moreover, it is notable that the \rsqcon values are nearly identical to the \rsqmar values in all three statistical models. This indicates that the \ri has only a minimal influence on the \rps value, thereby supporting the reproducibility of the benchmarking procedure.
\begin{table} [h]
\centering
\caption{Coefficients of determination for the three LMMs.} \label{tab:rsqrvalues}
\begin{tabular}{|c|c|c|}
\hline
\textbf{Statistical model} & \rsqconbf & \rsqmarbf \\
\hline
\lmmi & \lmmirsqcon & \lmmirsqmar \\
\lmmii & \lmmiirsqcon & \lmmiirsqmar \\
\lmmiii & \lmmiiirsqcon & \lmmiiirsqmar \\
\hline
\end{tabular}
\end{table}

\indent The influence of fixed effects and their interactions on the \rps value is estimated by examining the results of separate ANOVAs for the three LMMs\ifmasterthesis
\else
.
\fi
The analysis reveals that all fixed effects and their interactions significantly predict the \rps observations for all three statistical models, always with $p$-values below $0.0001$. Subsequently, the post-hoc tests are employed to delineate precisely which levels of the variables exhibit significant differences and, thus, exert a particular influence.
\subsubsection{Inference Engine Comparison}
\begin{figure}[h]
    \centering
    \includegraphics[width=0.49\textwidth]{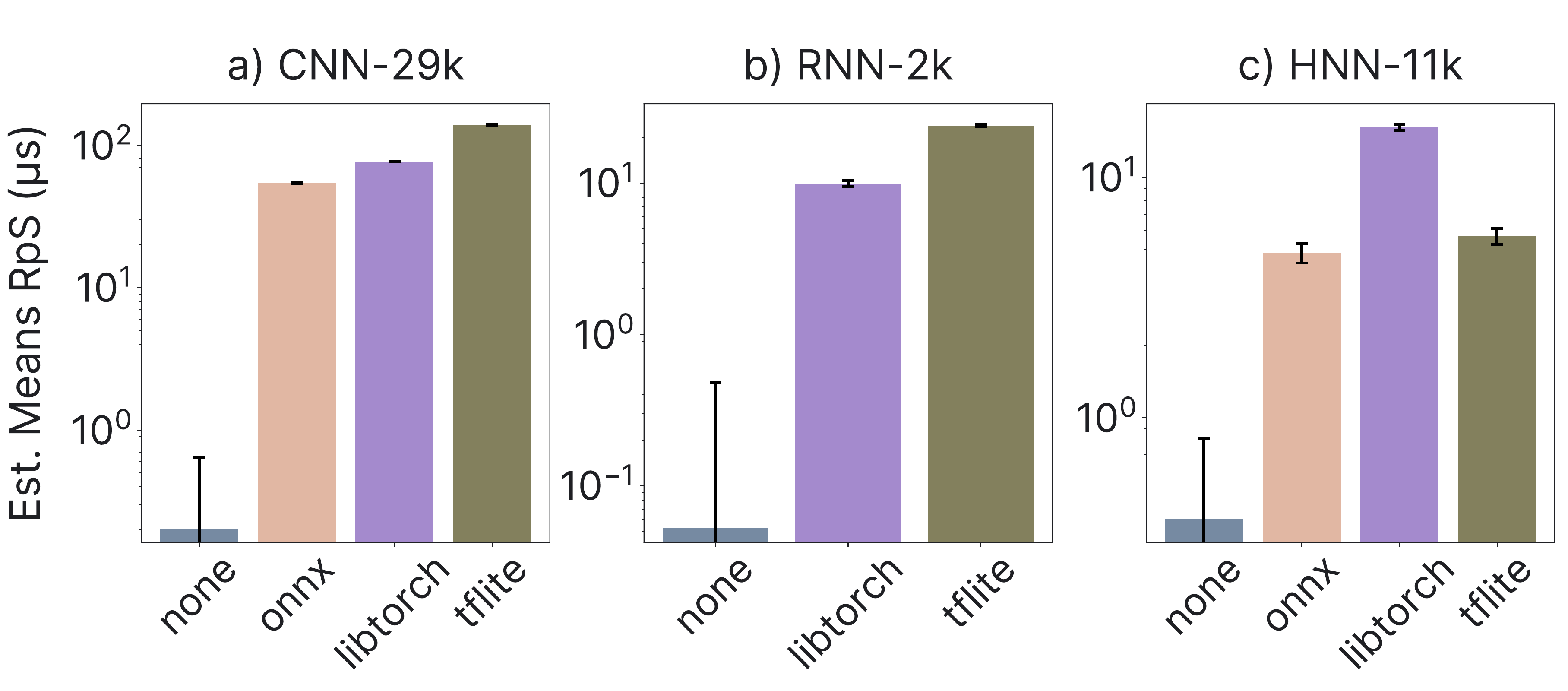}
    \caption{Estimated marginal means and 95\% CI of the \rps for the \ie predictor across different \mas. The values for Subfigure a and c have been computed from the \lmmi, while those for Subfigure b have been derived from \lmmii. The none legend denotes the \none.} \label{fig:ie-comparison}
\end{figure}

The first predictor that is subjected to a closer examination is the \ie. In both the \lmmi and \lmmii, all contrasts between the \ie variables are found to be highly significant ($p<0.0001$). The order from the fastest to the slowest \ie in the \lmmi is: \none, \onnx, \libtorch, \tflite. In the \lmmii, the contrasts show the same pattern, with \tflite, performing worse than \libtorch and the \none being the fastest.\\
\indent \Ma-wise comparisons of the \ie predictor for the \lcnn and \hybrid demonstrate that \onnx consistently outperforms \libtorch and \tflite. These comparisons are all highly significant ($p<0.0001$), with the exception of the difference between \onnx and \tflite for the \hybrid ($p=0.045$). The outcome for the \tflite versus \libtorch comparison exhibited less consistency. \tflite exhibits inferior performance compared to \libtorch for the \lcnn and \stateful, whereas the opposite is true for the \hybrid. The estimated marginal means and 95\% CI for the \ie and \ma combinations are depicted in Fig. \ref{fig:ie-comparison}.
\subsubsection{Influence of the Iteration}
\begin{figure}[h]
    \centering
    \includegraphics[width=0.49\textwidth]{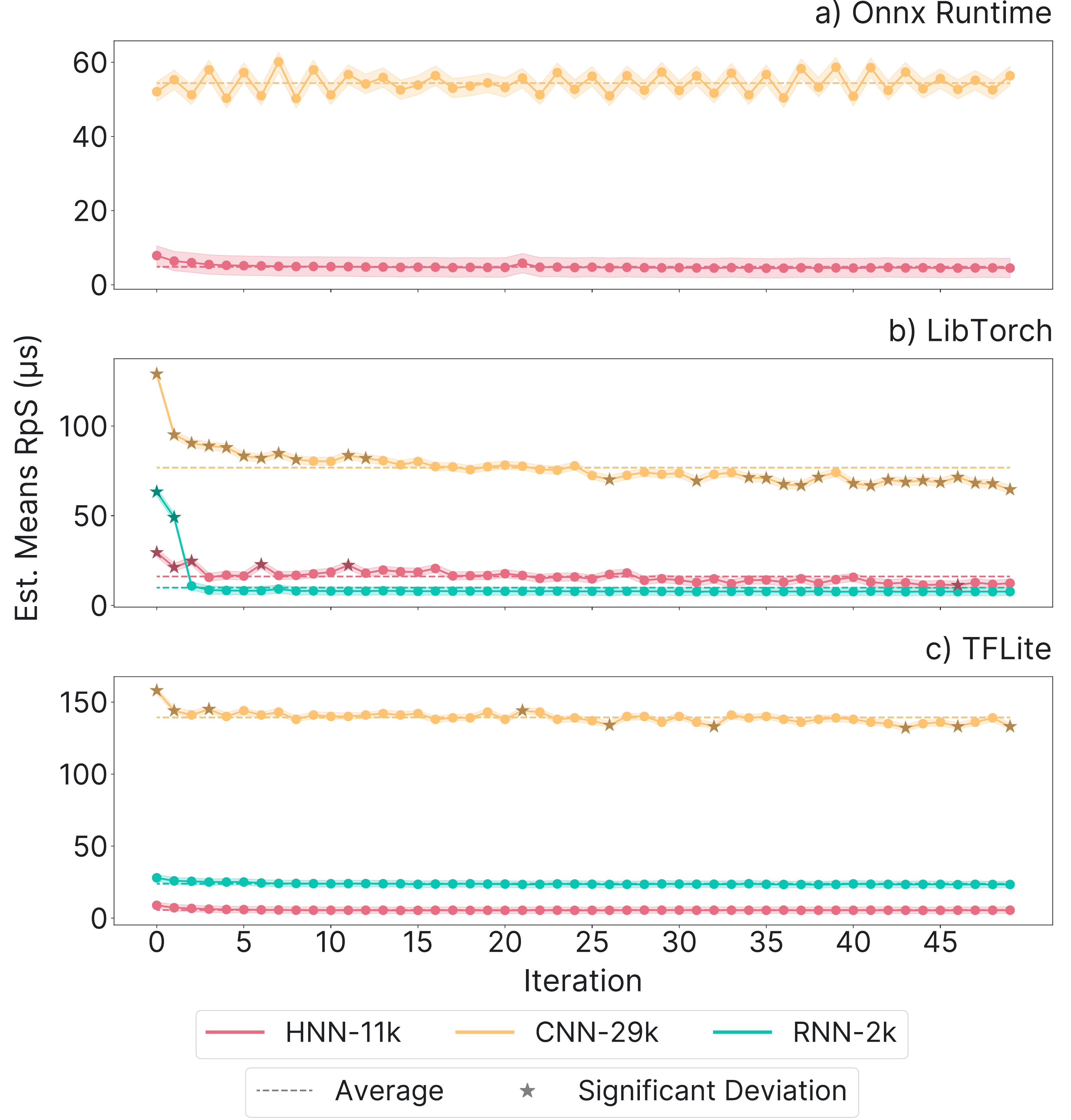}
    \caption{Estimated marginal means and 95\% CI of the \rps for different \ites, across different \ies and \mas. The dashed lines represent the average \rps, while the stars indicate highly significant differences compared to the average ($p < 0.0001$). For Subfigure a, the values are derived from \lmmi, while for Subfigure b and c, they are derived from \lmmii.} \label{fig:ite-comparison}
\end{figure}

In light of the varying numbers of real-time violati ons observed at different inference counts (cf. Section \ref{sec:results-radsan}), it becomes pertinent to investigate whether the \ite exerts an influence on the performance of the \ie for a specified \ma. The effect contrasts of the \ite predictor in the \lmmi and \lmmii reveal that this can be answered in the affirmative for certain \ie and \ma combinations.\\
\indent In particular, \libtorch exhibits significantly higher \rps values for the first \ites compared to the average \rps across all \mas ($p < 0.0001$). This effect is most pronounced for the \lcnn, where the first nine \ites exhibit significantly higher \rps values compared to the average. For \tflite, significantly higher \rps values are observed only for the \lcnn, but a tendency towards decreasing \rps values after the first iterations can also be observed for the two other \mas. A similar tendency is observed for \onnx inferencing the \hybrid. In contrast, the \lcnn with \onnx does not exhibit a tendency towards decreasing \rps values, but rather a tendency towards altering \rps values for consecutive \ites.\\
\indent The \none exhibits extremely high CI values, even crossing the zero line, indicating that the \none has been modeled in a way that makes its \rps impossible to discern from zero. Therefore, it is excluded from Fig. \ref{fig:ite-comparison}, which depicts the results of the effect contrasts by displaying the estimated marginal means, averages and CI for the \ite predictor \ie- and \ma-wise.
\subsubsection{Influence of the Buffer Size}
\begin{figure}[h]
    \centering
    \includegraphics[width=0.49\textwidth]{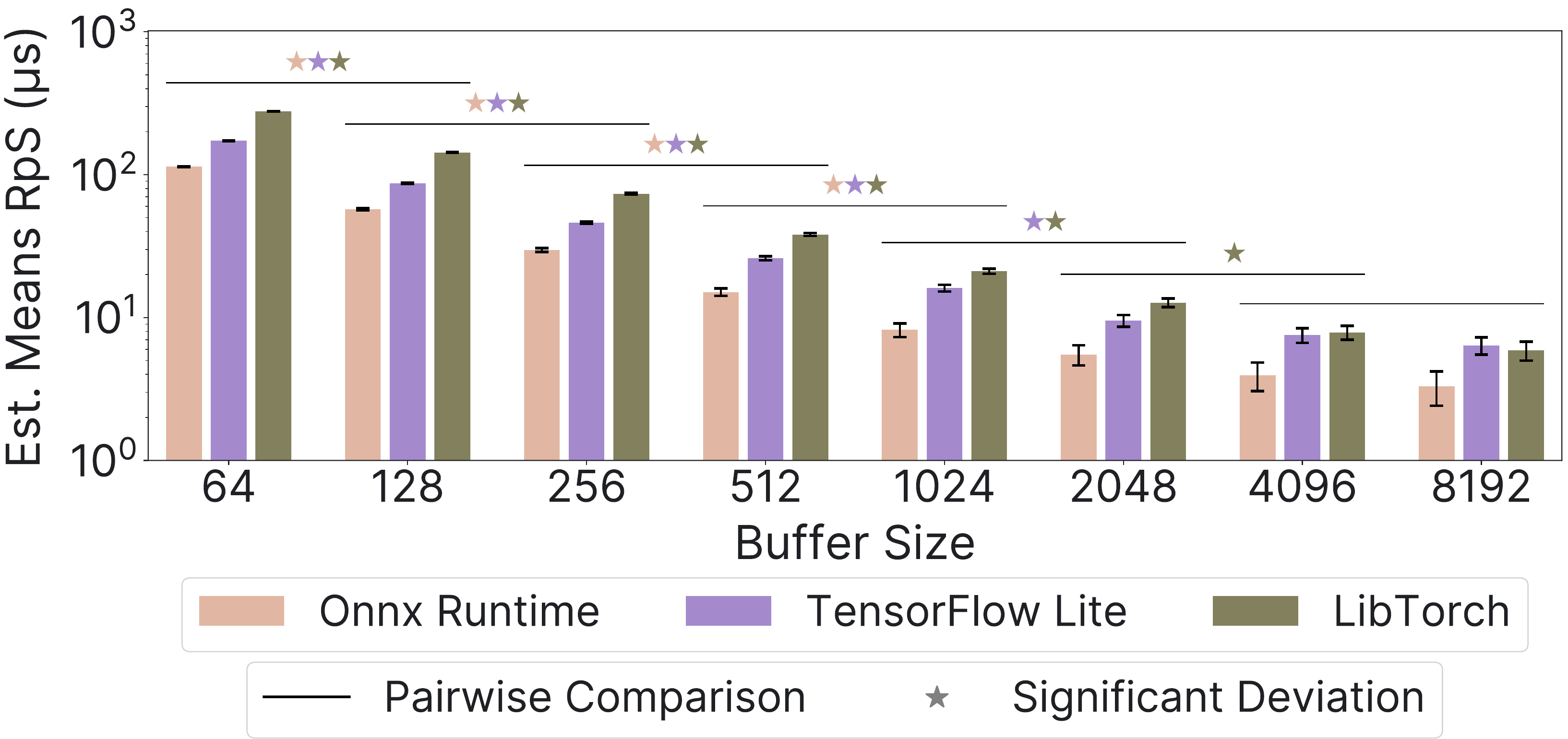}
    \caption{Estimated marginal means and 95\% CI of the \rps, calculated from the \lmmi for different \bss, across various \ies. Each \bs is compared pairwise to the next larger \bs of the same \ie, indicated by the horizontal line above, with stars denoting highly significant differences (p < 0.0001).} \label{fig:buffersize-comparison}
\end{figure}

The influence of the \bs on the \rps values is analyzed using the \lmmi. The results of the pairwise comparisons for the \bs predictor reveal that almost all \bs pair differences exert a highly significant influence ($p < 0.0001$) on the \rps values \ie-wise. Fig. \ref{fig:buffersize-comparison} depicts the estimated marginal means and CI for the \bs predictor. Due to its considerable large CI values, the \none is excluded from the analysis. The results indicate that the \rps values decrease with increasing \bs in a remarkably consistent manner across the three \ies. The impact of the \bs on the \rps values is particularly evident at low \bss and becomes less pronounced with increasing \bs.
\subsubsection{CNN Hyperparameter Comparison}
\begin{figure}[h]
    \centering
    \includegraphics[width=0.49\textwidth]{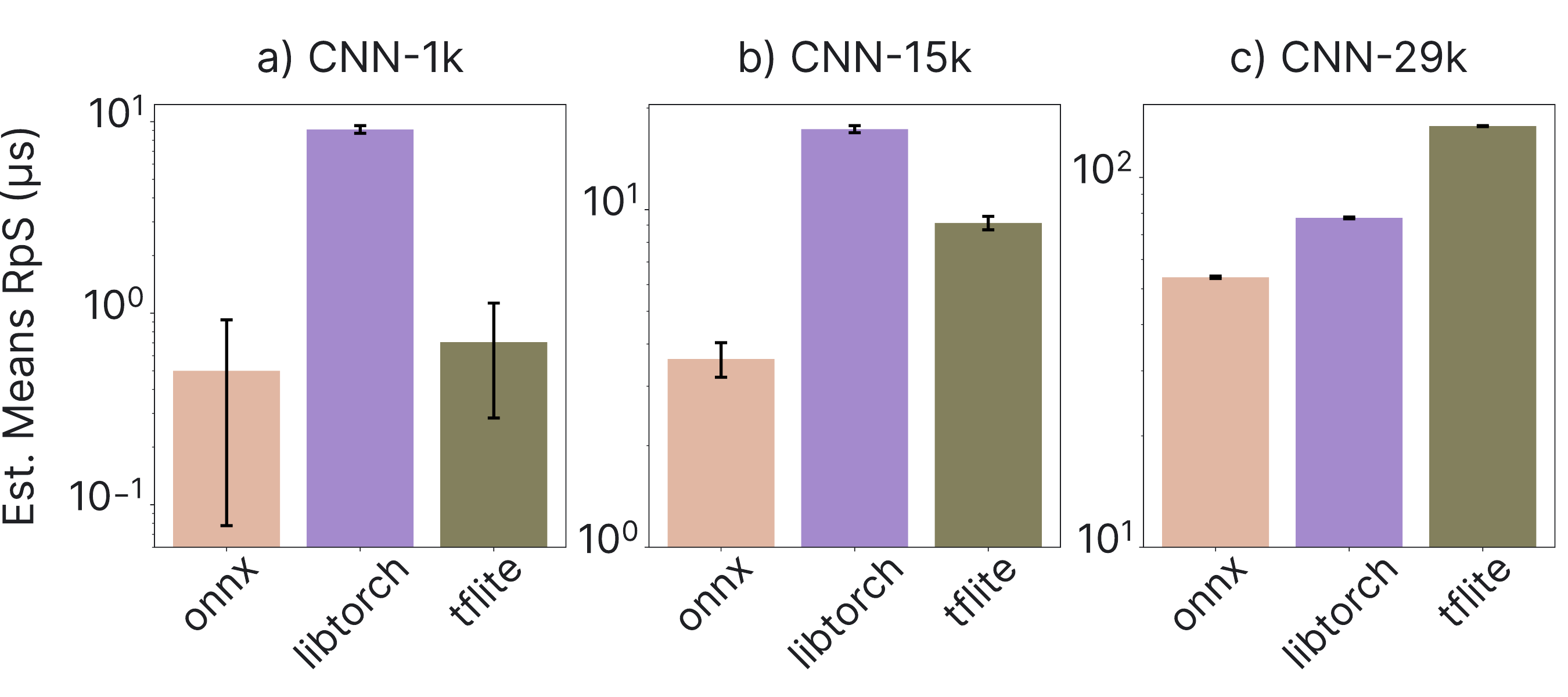}
    \caption{Estimated marginal means and 95\% CI of the \rps for the \ie predictor across different CNN \mas. The values have been computed from the \lmmiii.} \label{fig:cnn-comparison}
\end{figure}

In this follow-up study, the performance of the \ies across various hyperparameter combinations that define the size of the CNN architecture is evaluated. The \lmmiii is employed to conduct the required pairwise contrasts for the \ie predictor \ma-wise. The contrasts indicate that while \tflite exhibits significantly inferior performance compared to \libtorch for the \lcnn ($p < 0.0001$), a different picture emerges when the comparison is conducted at smaller sized \mas. As the size decreases, \tflite becomes increasingly faster than \libtorch. This phenomenon reaches a point where \tflite no longer exhibits significantly different \rps measures from \onnx at the smallest \ma (\scnn). In all other cases, \onnx outperforms \libtorch and \tflite with significant differences. The estimated marginal means and 95\% CI for all \ie and \ma combinations are depicted in Fig. \ref{fig:cnn-comparison}.
\section{Discussion} \label{sec:discussion}
This work presents \anira, a novel cross-platform inference library designed to address the unique challenges of integrating neural network inference into real-time audio applications. The library supports \tflite, \onnx, and \libtorch, which are amongst the most commonly used inference engines. \Anira and the inference engines were evaluated in terms of real-time safety and performance with regard to inferring three distinct neural network architectures: CNN, RNN, and hybrid networks. In order to assess performance, the integrated benchmarking functionality of \anira was employed in conjunction with statistical modeling.\\
\indent As previously observed, the aforementioned inference engines are not entirely real-time safe \cite{chowdhury_rtneural_2021}, \cite{stefani_comparison_2022}. By quantifying the real-time violations we discovered their persistence across all engines and models. The \libtorch inference engine exhibited the highest number of violations, likely due to its flexible design. In terms of model architecture, the stateful RNN, possibly due to its internal state, required high amounts of memory allocation. Furthermore, the majority of real-time violation tests demonstrated an increase in occurrences during the initial inferences, although not exclusively. This finding is at contrast to that presented in \cite{stefani_comparison_2022}, which indicated that all engines consistently ensure real-time safety after the initial inference. This discrepancy may be attributed to the more complex network architectures that were analyzed in this work and also to the different real-time violation detection method that was employed. While we employed the code sanitizer \radsan \cite{serebryany_addresssanitizer_2012}, which detects real-time violations by intercepting system library calls such as \codeword{malloc} and \codeword{pthread_mutex_lock}, the authors of \cite{stefani_comparison_2022} relied on monitoring the status of a hard real-time kernel, Xenomai Cobalt.\\
\indent To circumvent real-time violations, the \anira library employs a static thread pool to separate the inference from the audio callback. The static thread pool ensures that no potential oversubscription occurs, while its multiple threads allow for parallel inference of incoming data. The evaluation of the \anira library confirmed its robustness for real-time applications, as it did not exhibit any real-time violations within the audio callback and introduced only minimal runtime overhead per sample. Two different implementations for data sharing and synchronization between the audio callback and the inference threads are provided in \anira. The choice is left to the user, who may select either an atomic-based or a semaphore-based  approach. The atomic-based approach ensures that no system calls are made, whereas the semaphore-based approach accepts minor system calls, contingent on the operating system implementation, in exchange for a further reduction in latency.\\
\indent The integrated benchmarking capability for evaluating the runtimes of neural networks may be of particular interest for the deployment of audio applications, as it allows for the estimation of the maximum inference time. This measure is an important indicator of the real-time suitability of the network architecture and is a prerequisite for the built-in calculation of latency in the \anira library. Moreover, the standardized and user-friendly benchmarks have the objective of establishing comparability among neural network inference runtime measurements in real-time audio scenarios. We demonstrated that benchmarks with \anira can be run for various configurations and created three distinct datasets of runtime measurements.\\
\indent The statistical modeling of the datasets enabled the explanation of a high proportion of the variations in the runtime observations. The influence of the repetition index as random intercept, which accounts for fluctuations across different time windows in the benchmarks, was found to be minimal. Fluctuations may be attributed to various factors, including process scheduling, since our operating systems were not primarily designed for real-time purposes, allowing other processes to run in the background. Therefore, we conclude that the thread prioritization within the \anira library is functioning effectively.\\
\indent Post-hoc tests were employed to identify which levels of the variables exhibited significant differences and their respective impacts. Consequently, it was determined that \onnx is the fastest engine among all stateless models. When including stateful models in the comparison, which excluded \onnx, it was found that \libtorch outperforms \tflite on average. Furthermore, our investigation revealed significantly longer runtimes in early inferences, particularly for the model and inference engine combinations, which exhibited more frequent real-time violations during the initial inferences. Therefore, we recommend warm up phases before the audio callback to ensure more reliable runtimes. Additionally, when investigating the impact of different model input sizes on per-sample performance, we found that larger model input sizes lead to substantial performance gains. Consequently, for applications with less stringent latency requirements, the use of a larger model input size is advisable.\\
\indent While \onnx consistently outperforms the other two inference engines across all model architectures, the results per model were not consistent between \tflite and \libtorch. To identify the reasons for this inconsistency, a follow-up study was conducted, in which CNN models with different sizes were compared. The evaluation revealed that \tflite is faster for smaller architectures, but as the model parameters increase, \libtorch becomes faster. \onnx consistently remains the fastest inference engine overall in this comparison.\\
\indent Although the benchmarks provided valuable insights, there are limitations to consider. The architecture's performance was not evaluated in the context of model input size and host buffer size mismatches. While this approach allows for a more detailed analysis of the influence of the model input size, it does not reflect the circumstances of most audio applications, which typically have varying host buffer sizes. Furthermore, parallel inferences leveraging the thread pool were not benchmarked, which may have limited the assessment of performance.\\
\indent It is our hope that \anira will facilitate the integration of neural networks into real-time audio applications. Its flexibility and efficiency make it suitable for a diverse array of real-world applications, including audio plugins, speech processing, embedded audio systems, and smart musical instruments.
\section*{Acknowledgment}
For their invaluable support and guidance throughout the project, we would like to thank Prof. Dr. Stefan Weinzierl, Prof. Dr. Henrik von Coler, and Dr. Steffen Lepa.\\
\indent Furthermore, we acknowledge the use of the AI tools ChatGPT, DeepL Write and Github Copilot that have been employed for orthography, punctuation and grammar correction.
\printbibliography
\clearpage

\end{document}